\documentstyle[psfig,prd,aps,preprint]{revtex}
\begin{document}
\draft
\baselineskip 16pt
\renewcommand{\textwidth}{15cm}
\renewcommand{\topmargin}{.5cm}
\setlength{\textheight}{50pc}
\preprint{YITP-98-23,\ SU-ITP 98/18}
%\date{April, 1998}
\tighten
\title{ Cosmological constraints on primordial black holes produced
in the near-critical gravitational collapse %
}

\author{ Jun'ichi YOKOYAMA}
\address{\hfill\\
Department of Physics, Stanford University, Stanford, CA 94305-4060 and\\
Yukawa Institute for Theoretical Physics, Kyoto University, Kyoto
606-01, Japan\\
%\hfill\\
}

\maketitle

\abstract{The mass function of primordial black holes
created through the near-critical gravitational collapse is calculated
in a manner fairly independent of the statistical distribution of
underlying density fluctuation,
assuming that it has a sharp
peak on a specific scale.  Comparing it with various cosmological constraints
on their mass spectrum, some newly excluded range  is found in the
 volume fraction of the region collapsing into black holes
as a function of the horizon mass.
}


\maketitle

\newpage
%
%\def\theequation{\arabic{section}.\arabic{equation}}

        %%%%%%%%        Contents starts here  %%%%%%%%%%%%%%
\newcommand{\dw}{{\rm DW}}
\newcommand{\cw}{{\rm CW}}
\newcommand{\ml}{{\rm ML}}
\newcommand{\lt}{\tilde{\lambda}}
\newcommand{\lh}{\hat{\lambda}}
\newcommand{\phidot}{\dot{\phi}}
\newcommand{\phicl}{\phi_{cl}}
\newcommand{\adot}{\dot{a}}
\newcommand{\phat}{\hat{\phi}}
\newcommand{\ahat}{\hat{a}}
\newcommand{\hhat}{\hat{h}}
\newcommand{\phihat}{\hat{\phi}}
\newcommand{\Nhat}{\hat{N}}
\newcommand{\hth}{h_{th}}
\newcommand{\hbh}{h_{bh}}
\newcommand{\gsim}{\gtrsim}
\newcommand{\lsim}{\lesssim}
\newcommand{\bfx}{{\bf x}}
\newcommand{\bfy}{{\bf y}}
\newcommand{\bfr}{{\bf r}}
\newcommand{\bfk}{{\bf k}}
\newcommand{\bkp}{{\bf k'}}
\newcommand{\order}{{\cal O}}
\newcommand{\beq}{\begin{equation}}
\newcommand{\eeq}{\end{equation}}
\newcommand{\beqa}{\begin{eqnarray}}
\newcommand{\eeqa}{\end{eqnarray}}
\newcommand{\mpl}{M_{Pl}}
\newcommand{\lmk}{\left(}
\newcommand{\rmk}{\right)}
\newcommand{\lkk}{\left[}
\newcommand{\rkk}{\right]}
\newcommand{\lnk}{\left\{}
\newcommand{\rnk}{\right\}}
\newcommand{\call}{{\cal L}}
\newcommand{\calr}{{\cal R}}
\newcommand{\half}{\frac{1}{2}}
\newcommand{\kc}{\kappa\chi}
\newcommand{\bkc}{\beta\kappa\chi}
\newcommand{\gkc}{\gamma\kappa\chi}
\newcommand{\gbkc}{(\gamma-\beta)\kappa\chi}
\newcommand{\dchi}{\delta\chi}
\newcommand{\dphi}{\delta\phi}
\newcommand{\dOmega}{\delta\Omega}
\newcommand{\Phibd}{\Phi_{\rm BD}}
\newcommand{\echi}{\epsilon_\chi}
\newcommand{\ephi}{\epsilon_\phi}
\newcommand{\Phihat}{\hat{\Phi}}
\newcommand{\Psihat}{\hat{\Psi}}
\newcommand{\that}{\hat{t}}
\newcommand{\Hhat}{\hat{H}}
\newcommand{\zk}{z_k}
\newcommand{\msolar}{M_\odot}
\newcommand{\mbh}{M_{\rm BH}}
\newcommand{\bh}{{\rm BH}}
\newcommand{\calf}{{\cal F}}
\newcommand{\gtilde}{{\tilde g}}

%\baselineskip 0.8cm
%\newpage

In the study of gravitational collapse with formation of a black
hole a critical phenomenon was discovered by Choptuik
\cite{Choptuik}, who performed  a series of numerical calculation
of the evolution of a spherically symmetric system with a minimally coupled
scalar field.  He showed that the resultant mass of the black hole,
if formed at all, scaled as
$   \mbh\propto (p - p_c)^\gamma$ with $  p \geq p_c$,
where $p$ is a control parameter of the system and $p_c$ is the
critical value above which a black hole forms.  The remarkable
feature of the above formula is that the power-law exponent was found to
be independent of the choice of the control parameter.  Since then
a number of authors investigated various systems with scale-free
matter ingredients and found essentially the same result with
the unique value of the power index, $\gamma\simeq 0.36$
\cite{AE,EC}.  An
analytic explanation has also been given by Koike et al. \cite{Koike}.
Although this is certainly an interesting subject of study in
general relativity,
its relevance to the black holes created in astrophysical or
cosmological setting had  not  been known.
Recently, however, Niemeyer and Jedamzik \cite{NJ} observed the
same type of the critical behavior in their new numerical
calculation of the formation of primordial black holes (PBHs)
\cite{Zel,Haw}
out of radiation fluid in the Friedmann-Robertson-Walker background.

PBHs are formed in the radiation dominated era when a perturbed
region enters the Hubble radius if the amplitude of the fluctuation
exceeds some critical value $\delta_c \sim 1/3$ \cite{Carr}, which
has been confirmed by numerical analysis done two decades ago
\cite{NNP,BH}. It has also been concluded that the resultant PBH has a mass
of order of the horizon mass at formation.  Consequently  all the
studies done so far to constrain the mass spectrum of PBHs are based on
the assumption that PBHs of a specific mass is created at a
specific epoch, namely, when the mass scale entered the Hubble
radius as long as we are concerned with those formed during
radiation domination \cite{Nov}.  The same assumption has also
been made in building models of inflation to realize formation of
PBHs on some specific mass scales \cite{INN,Ra,hy,JY,BP,Kawa,cn}.

According to Niemeyer and Jedamzik \cite{NJ}, however, PBHs are
not only produced with about the horizon mass but also with much
smaller masses with the scaling formula
\beq
  \mbh=K(\delta - \delta_c)^\gamma,   \label{2}
\eeq
where as the control parameter $\delta$ they chose
the additional mass
in the perturbed region in unit of the horizon mass when the
relevant scale entered the Hubble radius.  They investigated two
classes of configurations of the perturbed region and found a
universal power index of $\gamma \cong 0.35$ in agreement with
Evans and Coleman \cite{EC} and Koike etal. \cite{Koike}.
They have also confirmed it is  independent of the choice of
the control parameter \cite{NJ}, as it should be.

Although the scaling relation (\ref{2}) is expected
to be valid only in the immediate neighborhood of
$\delta_c$, most black holes are expected to form with such an initial value
of $\delta$, because it has generically a rapidly declining
probability distribution function (PDF) near $\delta=\delta_c$
and the probability to find $\delta \gg \delta_c$ is exponentially smaller.
Hence it is sensible to calculate the expected mass function of
PBHs using the formula (\ref{2}).
The above-mentioned
 feature of the PDF also allows us to estimate the mass function
fairly independent of the specific form of the PDF of primordial
density or curvature fluctuations as seen below.

The purpose of the present paper is to investigate the effect of
the near-critical gravitational collapse
on the cosmological constraints on the mass
spectrum of PBHs and on the model building of inflation to produce
PBHs.  As a result
we find that, thanks to the fact that $\gamma$ is relatively
small, we do not have to change the conventional view to the issue
drastically except that we find a new constraint on the fraction of
the space collapsing to PBHs when the horizon mass is
in the range between $4\times10^{14}$g and $6\times10^{16}$g.

In this paper we study the case of a primordial spectrum of
density fluctuation that is
sharply peaked on a single but arbitrary mass scale just as in the
initial configuration adopted in \cite{NJ}.  Several models have
been proposed to realize such a spectral shape in inflationary
cosmology \cite{INN,JY,BP,Kawa,cn}.  Even in the case with such a
simple spectrum, the configuration of a perturbed region has
functional degrees of freedom and one must in principle calculate
both the probability to realize each initial configuration and the
resultant mass of the black hole, if formed at all, in order to
calculate the mass function of PBHs.  In the language of the
near-critical collapse, there are many possible one-parameter families
of the initial data in
the configuration functional space to approach the critical surface
and we find different values of $\delta_c$ and $K$ for each
family with the exponent $\gamma$ being the only universal
quantity.  Hence we should first calculate the mass function of
PBHs when the initial configuration is changed
along each one-parameter family near the critical surface
and then integrate over such families
in the functional space together with their relative probability
to obtain the overall mass function,
which is a formidable task.  Fortunately, however, there is yet
another common feature in PBH formation, which is the fact that
the typical mass of the black hole is of order of the horizon mass
independent of the shape of the perturbed region
even in the presence of the critical phenomenon as shown
explicitly in \cite{NJ}.  We thus reduce the problem with
infinitely many degrees of freedom to a one-dimensional issue,
that is, we adopt an approximation that the system is governed by a single
parameter, $\delta$, characterizing the amplitude of fluctuations,
and that the most common black hole has the
horizon mass, $M_H$, when the peak of fluctuation entered the
Hubble radius at $t=t_H$.

Since the PDF of $\delta$, $P(\delta)$, is a steeply declining function
around $\delta_c$ we write it as
\beq
  P(\delta)d\delta =e^{-f(\delta)}d\delta,  \label{pdf}
\eeq
where $f(\delta)$ is a well-behaved function around $\delta \simeq
\delta_c$.  If $\delta$ is Gaussian distributed, as assumed in most
literature \cite{Carr,CGL,GL}, $f(\delta)$ is explicitly
given by
\beq
   f(\delta)=\ln(\sqrt{2\pi}\sigma)+\frac{\delta^2}{2\sigma^2},
   \label{Gauss}
\eeq
with $\sigma$ being the dispersion of $\delta$.

Then the probability, $\beta(M_H)$, that the relevant mass
scale has an above-threshold amplitude of fluctuations to collapse
 into a black hole as it enters the Hubble radius is given by
\beq
  \beta(M_H) = \int_{\delta_c}P(\delta)d\delta
  = \int_{\delta_c}e^{-f(\delta)}d\delta,
\eeq
which is also equal to the volume fraction of the region collapsing to a
black hole at $t=t_H$.
Due to the fact that $P(\delta)$ is a steeply decreasing function the
integral is sensitive to only its lower bound $\delta_c$.  Furthermore
we can Taylor expand $f(\delta)$ as
\beq
 f(\delta)=f(\delta_c)+f'(\delta_c)(\delta - \delta_c)+\cdots
  \equiv f_c + q(\delta-\delta_c)+\cdots,  \label{Taylor}
\eeq
to find
\beq
  \beta(M_H)\cong \frac{1}{q}e^{-f_c}.   \label{Beta}
\eeq
The above formula (\ref{Beta}) based on the approximation
(\ref{Taylor}) is applicable if $|f''(\delta_c)| \ll q^2$.
In the case of a Gaussian distribution this corresponds to
$\sigma \ll \delta_c$.  Putting $\delta_c=1/3$ and using (\ref{Gauss})
in (\ref{Beta}) we recover Carr's formula \cite{Carr},
$\beta\simeq\sigma \exp[-1/(18\sigma^2)]$.
If $P(\delta)$ is non-Gaussian, we must analyze case by case, but as long
as we consider the case with small enough $\beta(M_H)$ in the
cosmologically allowed range \cite{Nov} and use specific non-Gaussian
distributions reported in the literature \cite{BP,cn,Ivanov}, we may
justify (\ref{Taylor}).  Hence this simple approximation enjoys a wide
applicability.

Then the contribution of PBHs to the density parameter
 at formation $t_f \gsim t_H$, is given by
\beq
  \Omega_\bh (t_f) = \frac{1}{M_H}\int_{\delta_c}\mbh (\delta)
  P(\delta)d\delta = \frac{1}{M_H}\int_{0}^{\infty}
  MP[\delta(M)]\frac{1}{\gamma}\lmk \frac{M}{K}\rmk^{\frac{1}{\gamma}}
  \frac{dM}{M},
\eeq
with
$  \delta(M)=\delta_c +(M/K)^{\frac{1}{\gamma}}$.
Using  (\ref{pdf}) and (\ref{Taylor}) in the above integral, the
differential mass spectrum reads
\beq
  \frac{d\Omega_\bh (M,t_f)}{d\ln M}
  =\frac{M}{\gamma M_H}\lmk \frac{M}{K}\rmk^{\frac{1}{\gamma}}
  \exp\lkk -f_c-q\lmk \frac{M}{K}\rmk^{\frac{1}{\gamma}}\rkk,
\eeq
which has a peak at
$   M_{\rm max}\equiv K q^{-\gamma}(1+\gamma)^\gamma.$
Since the typical black hole mass is around $M_H$ as stated above,
let us identify $M_{\rm max}$ with $M_H$ following the line of thought
explained above.
We then find
\beqa
  \frac{d\Omega_\bh (M,t_f)}{d\ln M}
  &=& \beta(M_H)\lmk 1+\frac{1}{\gamma}\rmk
  \lmk\frac{M}{M_H}\rmk^{1+\frac{1}{\gamma}}
  \exp\lkk - (1+\gamma)\lmk\frac{M}{M_H}\rmk^{\frac{1}{\gamma}}\rkk,
  \label{delomega}  \\
  &\cong& 3.86\beta(M_H)\lmk\frac{M}{M_H}\rmk^{3.86}
  \exp\lkk - 1.35\lmk\frac{M}{M_H}\rmk^{2.86}\rkk, \nonumber
\eeqa
which is now independent of $q$ and $K$.
Formal integration of (\ref{delomega}) from $M=0$ to $\infty$ yields
$
  \Omega_\bh (t_f) \simeq
  (1+\gamma)^{-1}\Gamma(\gamma)\beta(M_H)
  =0.80\beta(M_H) $
for $\gamma=0.35$, as opposed to the case all the PBHs have the horizon
mass where we would find $\Omega_\bh(t_f)=\beta(M_H)$.
Thus the total mass density of PBHs does not change practically
even if we take the effect of the near-critical collapse into
account.

The mass spectrum (\ref{delomega}) is depicted in Fig.\ 1 where we
find that the abundance of smaller-mass black holes are suppressed
even in the presence of the critical behavior. We can hence
convince ourselves that the previous assumption that only horizon-mass
black holes are likely be produced is a good approximation in model
building where we try to identify some astrophysical objects such
as MACHOs \cite{macho} with PBHs \cite{JY,Kawa,cn}.
Indeed, if too many Jupiter-mass PBHs were produced simultaneously
in an attempt to produce MACHO-mass PBHs, PBH-explanation of
MACHOs would be ruled out because the observation of the MACHO group
is very
sensitive to Jupiter-mass objects and their abundance in the
galactic halo has already been stringently constrained \cite{M2}.
However, the mass fraction (\ref{delomega}) implies we are free
from such a problem.

\begin{figure}[htb]
  \begin{center}
%    \epsfile{file=potential.eps,width=10cm}
  \leavevmode\psfig{figure=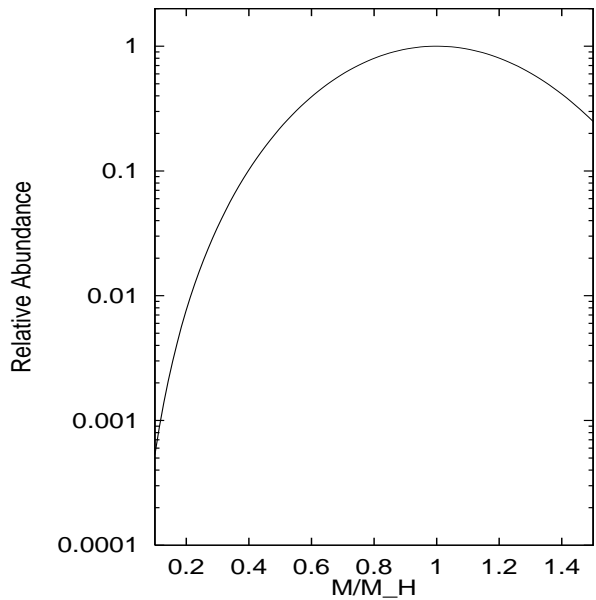,width=8.5cm,height=8cm}
  \end{center}
  \caption{Expected mass function of PBHs (\ref{delomega})
 created through the
near-critical gravitational collapse in the case density
fluctuation has a sharp peak on the mass scale $M_H$.}
  \label{fig:1}
\end{figure}

We must, on the other hand,
consider its effect on cosmological constraints more carefully.
At later time $t$ we find the mass spectrum, $\calf (M,t;t_f)$, of
the holes produced at $t_f$ as
\beqa
  \calf (M,t;t_f) \equiv \frac{d\Omega_\bh (M,t;t_f)}{d\ln M}
  &=& \frac{d\Omega_\bh (M,t_f)}{d\ln
  M}\lmk\frac{t}{t_f}\rmk^{\frac{1}{2}},~~~~~t<t_{eq},  \nonumber \\
  &=& \frac{d\Omega_\bh (M,t_f)}{d\ln
  M}\lmk\frac{t_{eq}}{t_f}\rmk^{\frac{1}{2}},~~~~~t>t_{eq},
  \label{ft}
\eeqa
where
\beq
t_{eq}=4.2\times 10^{10}(\Omega_0h^2)^{-2} {\rm sec}
=1.9\times 10^{12}\Gamma^{-2} {\rm sec},~~~~\Gamma\equiv
\frac{\Omega_0}{0.3}\lmk\frac{h}{0.7}\rmk^2=\frac{\Omega_0h^2}{0.147},
\eeq
is the equality time with $\Omega_0$ and $h$ being current values of
the total density parameter and the Hubble parameter in unit of
100km/sec/Mpc, respectively.

Cosmological constraints on the mass spectrum of the PBHs can be
classified to three classes.  The first one applies to heavy black
holes with mass $M_\bh > 4\times 10^{14}$g which have not
evaporated by now.  Current mass density of such holes should not
exceed the total mass density of the universe.
The second class is due to the
radiation of high energy particles from evaporating black
holes \cite{evaporate,carre}.
Various constraints have been imposed from primordial
nucleosynthesis \cite{nuc}, microwave background radiation
\cite{cmb}, and gamma-ray background radiation \cite{gamma}.
Finally, there may be yet another class of constraints if PBHs do not
evaporate completely but leave relics with mass of order of the
Planck mass
or larger \cite{relic}.  Their mass density should remain small
enough.

Even if we adopt the new mass spectrum (\ref{2}), cosmological
constraints of the first and the third classes described above are
not altered practically because the total number of black holes as
well as the abundance of most typical black holes, which have
approximately the horizon mass at formation and
dominates their mass density, are not expected to change
significantly as shown above.  In particular, the constraint of
the first class, namely, that from the total mass density of PBHs reads
\beq
  \beta (M_H) < 2\times 10^{-19}\Gamma\lmk\frac{M_H}{4\times
  10^{14} {\rm g}}\rmk^{\frac{1}{2}}\frac{\Omega_0}{0.3},
   \label{massdensity}
\eeq
demanding that $\Omega_\bh$ today should not exceed $\Omega_0$.

We now concentrate on the constraints associated with
evaporation.  The lifetime of a black hole with mass $M$
is given by
\beq
  t_{ev}(M)=1.7\times10^3\gtilde^{-1}\lmk\frac{M}{\mpl}\rmk^3t_{Pl},
  \label{evaptime}
\eeq
where $\mpl$ $(t_{Pl})$ is the Planck mass (time) and
$\gtilde$ is the effective number of massless state radiated
from the black hole in units of 7.25 which applies at the
low-energy or large mass limit \cite{GL}.
A black hole will evaporate before the equality time if its mass
satisfies the inequality
$ M < 6\times10^{12}\gtilde^{\frac{1}{3}}
  \Gamma^{-\frac{2}{3}}{\rm g}\equiv M_{eq}$.
Eliminating $t_f$ from both sides of (\ref{ft}) by virtue of
the relation $M_H=\mpl^2t_H\simeq \mpl^2 t_f$, we find
\beqa
  \calf (M,t_{ev}(M);M_H)= 1.6\times10^2\gtilde^{-\frac{1}{2}}
  \beta (M_H)\lmk\frac{M}{M_H}\rmk^{4.36}\frac{M}{\mpl}
  \exp\lkk - 1.35\lmk\frac{M}{M_H}\rmk^{2.86}\rkk,&~&M<M_{eq} \label{massfn}\\
  = 2.3\times10^{28}\Gamma^{-1}\beta (M_H)
  \lmk\frac{M}{M_H}\rmk^{3.86}\lmk\frac{\mpl}{M_H}\rmk^{\frac{1}{2}}
  \exp\lkk - 1.35\lmk\frac{M}{M_H}\rmk^{2.86}\rkk,&~&M>M_{eq}.
  \nonumber
\eeqa

The above mass spectrum should be compared with the following
constraints of the second class \cite{Nov,CGL,GL,carre,nuc,cmb,gamma}.
\beqa
 \calf (M,t_{ev}(M)) \lsim&
 1\times10^{-2}M_{10}^{\frac{1}{2}},~~~&{\rm for}~
 M=10^9-10^{10}{\rm g},\nonumber \\
 \lsim&5\times 10^{-7}M_{10}^{\frac{3}{2}},~~~&{\rm for}~
 M=10^{10}-10^{11}{\rm g},\nonumber \\
 \lsim&2\times 10^{-8}M_{10}^{\frac{7}{2}},~~~&{\rm for}~
 M=10^{11}-6\times10^{11}{\rm g},\label{constraint} \\
 \lsim&4\times10^{-2},~~~&{\rm for}~
 M=6\times10^{11}-10^{13}{\rm g},\nonumber \\
 \lsim&1\times10^{-8},~~~&{\rm for}~
 M\simeq 4\times10^{14}{\rm g},\nonumber
\eeqa
where $M_{10}\equiv M/10^{10}$ g.

Putting $M=M_H$ in (\ref{massfn}) and comparing it with
(\ref{constraint}), we practically recover the constraints on $\beta(M_H)$
previously obtained in the literature, which are depicted by a solid
line in Fig.\ 2.  In the presence of the near-critical collapse,
however, we should also consider constraints on $\beta(M_H)$
arising from PBHs with mass $M \ll M_H$.  In fact, however, due to
the steep shape of (\ref{massfn}) we find a newly excluded range
\footnote{There may be another small newly excluded region around $M_H \gsim
10^{11}$g.  However, we did not depict it in Fig.\ 2
since it is based on a rather qualitative constraint from
nucleosynthesis obtained two decades ago, and refined calculations may
well alter it.}
only in the region
$M_H > 4\times10^{14}$g, which is imposed from the last constraint of
(\ref{constraint}), namely,
\beq
  \beta(M_H)< 2\times 10^{-27}\Gamma\lmk\frac{M_H}{4\times
  10^{14}{\rm g}}\rmk^{4.36}\exp\lkk 1.35\lmk\frac{M_H}{4\times
  10^{14}{\rm g}}\rmk^{-2.86}\rkk.
\eeq
\begin{figure}[htb]
  \begin{center}
%    \epsfile{file=potential.eps,width=10cm}
  \leavevmode\psfig{figure=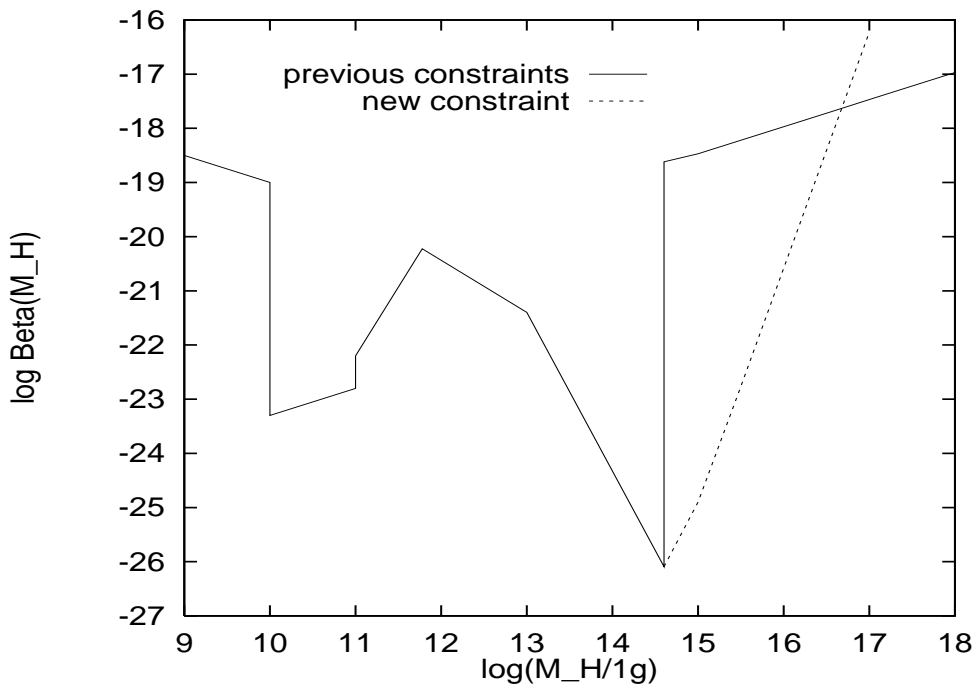,width=13.5cm,height=9cm}
  \end{center}
  \caption{Cosmological constraints on the volume fraction of
the region, $\beta$, above the threshold of PBH formation when the
horizon mass is equal to $M_H$.  The region between the solid and
 the dotted lines
is newly excluded.}
  \label{fig:2}
\end{figure}
\noindent
As is seen in Fig.\ 2, this new constraint is more
stringent than (\ref{massdensity}) for
$M_H < 6\times 10^{16}$g.

It has been argued that
high energy phenomenon associated with currently evaporating
black holes with mass $\sim 4\times 10^{14}$g can explain the origin of a
class of gamma-ray burst \cite{Cline}.  If this is the case,
their  abundance should be around $\Omega_\bh=10^{-8}$ today.
If such a tiny amount of black
holes were created at the low-mass tail of the near-critical collapse,
one could explain the origin of both such bursts and dark matter
simultaneously.  That is,
PBHs with mass around $6\times 10^{16}$g could be the dominant dark matter
component which makes up $\Omega \simeq 0.3$ today.

In summary, we have obtained an approximate but generic form of
the mass function of primordial black holes which are produced
through near-critical gravitational collapse of radiation fluid in
the early universe in the case density fluctuations have a sharp
peak on a specific scale. It is fairly independent of
the statistical distribution of density fluctuations.
Due to the smallness of the critical
exponent $\gamma=0.35$ the resultant mass function has a steep
spectrum in the low-mass tail,
so that the previous assumption that the PBHs are
created with nearly the horizon mass at formation is basically
correct.  Nevertheless we find some newly excluded range in the
 volume fraction of the region with $\delta  > \delta_c$
as a function of the horizon mass as depicted in Fig.\ 2.

\vspace{1cm}

The author is grateful to Professor Andrei Linde for
his hospitality at Stanford University, where this work was done.
This work was partially supported by the Monbusho.

\end{document}